# Plasmono-Atomic Interactions on a Fiber Tip

Eng Aik Chan[1*], Giorgio Adamo[1*], Syed Abdullah Aljunid[1], Martial Ducloy[1,3], Nikolay Zheludev[1,2], and David Wilkowski[1,4,5]

[1]Centre for Disruptive Photonic Technologies, TPI & SPMS, Nanyang Technological University, 637371, Singapore, Singapore.

[2]Optoelectronics Research Centre and Centre for Photonic Metamaterials, University of Southampton, Southampton SO17 1BJ, UK

[3]Laboratoire de Physique des Lasers, UMR 7538 du CNRS, Université Paris13-Sorbonne-Paris-Cité, F-93430, Villetaneuse, France.

[4]MajuLab, International Joint Research Unit UMI 3654, CNRS, Université Côte d'Azur, Sorbonne Université, National University of Singapore, Nanyang Technological University, Singapore, Singapore.

[5]Centre for Quantum Technologies, National University of Singapore, 117543, Singapore, Singapore.

*Author e-mail address: echan003@e.ntu.edu.sg*

**Abstract:** Light-atom interaction can be engineered by interfacing atoms with various specially designed media and optical fibers are convenient platforms for realization of compact interfaces. Here, we show that an optical fiber sensor bearing a plasmonic metasurface at its tip can be used to detect modifications of the Doppler-free hyperfine atomic spectra induced by coupling between atomic and plasmonic excitations. We observed the inversion of the phase modulation reflectivity spectra of Cesium vapor in presence of the metamaterial. This work paves the way for future compact hybrid atomic devices with a cleaved tip as substrate platform to host various two-dimensional materials.

Atomic vapors are commonly used for frequency standards, metrology, quantum technologies and for high-precision tests in fundamental physics [1-4]. However, optical experiments with atomic vapors are generally undertaken with bulky cells and free-space optical setups. Incorporation of atomic media with fiber technology would considerably miniaturize the platform. Important steps towards that direction have been done using hollow core fibers [5], fiber integrated-capillary devices [6], waveguides [7,8], nanofibers [9,10], filled up with atomic or molecular gases for compact spectroscopic applications. In this letter, we propose an alternative solution using a cleaved tip of a multimode optical fiber immersed into an atomic vapor. A decisive advantage of our setup is the fact that it offers a simple substrate platform (the cleaved tip) to host various materials such as 2D materials stack, hyperbolic and topological metamaterials, which can be used to engineer the properties of the atom-light interactions. Here, we illustrate this scheme by using a metallic nano-structured metasurface, whose surface plasmon resonances are tunable in wavelength by design, which leads to Fano-like couplings [11,12], and modifications of the Casimir-Polder potential [13-15].

The spectroscopic measurement is performed using selective reflection technique at the cleaved tip of the fiber [16,17]. By construction, the excitation laser field is at near-normal incidence at the fiber tip interface, leading to a Doppler-free and surface-sensitive -within a probing layer of $1/k$ distance- spectroscopic signal [18], where $k$ is the wavenumber of the optical beam. This gives the opportunity to access the natural spectral line, in a single laser beam experiment, by measuring the reflection through the optical fiber tip immersed in an atomic vapor. As the variation of the interface reflectance, induced by the presence of vapor, is small, we used a sensitive phase modulation spectroscopy technique to recover the sharp atomic line signature from the broad reflectance spectrum of the interface. We implemented this technique by phase modulating a narrowband laser and demodulating the reflection signal at the frequency of modulation. By scanning the main frequency of the laser source across the atomic resonance it is possible to recover the spectrum of atomic vapor at the vicinity of the fiber tip interface.

The optical set-up is presented in Figure 1(a). The laser radiation is produced by a diffraction-grating tunable semiconductor diode laser (Toptica DL pro) operating on the 852 nm $6^2S_{1/2}$ - $6^2P_{3/2}$ D2 line of Cesium [see Figure 1(b)]. The beam is phase-modulated at a frequency of 9 MHz by an electro-optic modulator before being launched into an optical fiber. We used a multimode silica fiber (Thorlabs FG200UEA) with core diameter of 200 µm and a numerical aperture (NA) of 0.22 (acceptance angle of 13°). The large core of the optical fiber was required to sustain a significant level of light power at the fiber tip, without entering nonlinear regimes of interactions with the atomic vapor (saturation intensity 2.2 mW/cm$^2$). The optical fiber tip is immersed in an ultrahigh vacuum chamber and sealed on a Swagelok connector with a Teflon ferrule[19]. Cesium vapor was introduced into the chamber, and maintained at a thermal equilibrium of 50°C. Optical transmission spectroscopy through the chamber's viewports was done and the Cesium vapor density is determined to be $5 \times 10^{11}$ cm$^{-3}$ in agreement with the saturated vapor pressure at the temperature of the chamber. The reflected light intensity from the fiber tip is detected by a silicon avalanche photodetector (Hamamatsu C10508) and demodulated using a sensitive lock-in amplifier (Zurich instruments HF2LI). A part of the laser radiation is diverted to a hole-burning set-up with a room-temperature Cesium cell for frequency calibration.

To account for the modifications induced by the nanostructure on the atomic spectra, we compare the Cesium vapor reflection spectrum at a bare fiber tip with the Cesium vapor reflection spectrum at the interface with a plasmonic metasurface with surface plasmon resonance overlapping with the D2 line of Cesium. The metasurface is prepared by depositing a 50 nm thick layer of silver followed by a 3 nm capping layer of silicon dioxide on the tip of a cleaved fiber facet by thermal evaporation. Then, we structure the layers with a two-dimensional array of rings, via Focused Ion Beam (FIB) milling. The rings are designed to have a 120 nm inner diameter, a 190 nm outer diameter, with a 35 nm width, a 50 nm depth, and are arranged on a square lattice with period of 370 nm along both directions (see Figure 1(c) and inset of Figure 1(a)). We ensure that the nanostructures were completely carved through the silver layer. The metasurface has a 50 THz wide surface plasmon resonance centered at a wavelength of 830 nm, thus slightly blue shifted with respect to the 852 nm Cesium D2 line. The experimental reflectance spectra of the metasurface, for two orthogonal linear polarizations are shown in Figure 2(a). Full-wave finite element method electromagnetic simulated spectra Figure 2(b), by COMSOL Multiphysics, show a fair agreement with the experimental ones for both polarizations. The positions of the two resonant dips match well and the discrepancy in modulation depth and width could be attributed to the losses introduced by FIB milling, due to ~~Ga~~ contamination from Ga ion source during milling, fabrication imperfections, and angle dependence of the plasmonic resonance position [20,21]. Since the polarization of the light is not preserved in the large core multimode fiber, it is important that the reflection curves are independent of the polarization, as shown in Figures 2(a) and (b). This result is a consequence of the axial symmetry of our fiber-metamaterial system. Moreover, the incident angle of the light field on the metamaterial is not properly defined but limited to the NA of the fiber. For such a moderate angle range up to 13°, we did not expect polarization-angle dependence of the plasmonic resonance, as it was previously studied with similar arrangements (see for example ref. [20]).

Figures 2 (c) and (d) show a comparison of the demodulated reflective spectra recorded at the bare fiber tip and at the fiber tip equipped with the plasmonic metasurface. The larger noise level, observed for the metasurface fiber, is due to the lower light transmission of the metasurface, with respect to the bare fiber case. One important feature of these spectra is the inversion of the dispersion-like shape upon transition from bare fiber to metasurface fiber. Such effect has been previously observed using a free-space optical setup probing on atomic gases hybridized with highly confined plasmons [22,23]. In those works it was shown that the modifications in line shape, as observed in Figures 2 (c) and (d), are associated with Fano-like coupling between the narrow atomic resonance and the plasmonic resonance whose linewidth is several orders of magnitude broader than the atomic resonance.

Beside Fano-like coupling, the shape of Doppler-free reflective spectra are also modified by the Casimir-Polder potential exerted by a surface on the atoms [18]. Using a classical picture, the Casimir-Polder interaction can be understood as the coupling of an atomic dipole moment with its own image reflected by a surface [24]. When an atom is located above a perfect reflective surface, the latter induces an anti-correlated dipole image leading to an attractive potential, which is the Casimir-Polder potential when propagation effects are taken into consideration [13]. If the perfect surface is replaced by a dispersive surface exhibiting a surface mode that is in near-resonance with the atomic transition, the induced image of the mean dipole moment is both enhanced and de-phased, resulting in either an enhanced attractive or a repulsive Casimir-Polder potential of the excited state [25,26]. We note that the Casimir-Polder potential of the atom ground state is solely due to the fluctuations of the dipole moment that are almost insensitive to the surface mode resonance [27]. The plasmonic metasurface equipped on our fiber tip provides a resonant surface mode that modifies the Casimir-Polder potential. To evaluate the modification of the Casimir-Polder potential on the fiber tip, we performed a fit of the demodulated reflectance spectra recorded at the bare fiber tip and the fiber tip coated with plasmonic metasurface, following the theoretical treatment in ref. [27]. The fitting parameters are the atomic linewidth $\Gamma$, the differential Van der Waals (Casimir-Polder in the non-retarded regime) coefficient $\Delta C_3$ between the atomic excited and ground states and an overall amplitude $V_0$. The fits are shown as black dashed lines in Figures 2(c) and (d). The Van der Waals coefficient $\Delta C_3$ obtained from the fit is 3(2) kHz µm$^3$ for the bare fiber and 6(6) + i5(7) kHz µm$^3$ for the metasurface-coated fiber. Similar enhancements of the Van der Waals coefficients on Cesium vapor coupled with blue-detuned plasmonic metasurfaces on a dielectric window and reduction of the Casimir-Polder interaction on Cesium vapor coupled with red-detuned plasmonic metasurfaces were reported [27]. We note that the theoretical treatment, to derive the lineshape function of the atomic gas at the metamaterial interface used in the fits, is carried out in the mean-field approximation. The latter consists of replacing the metasurface by an effective bulk material having the same thickness and the same far field properties at normal incidence. From the far field properties, we derive the complex refraction index of the effective bulk material, which, in our case, is 0.67 + 2.53i [28,29]. At a first sight, it seems surprising that the mean field approximation correctly describes the selective reflection spectroscopy signal of the atomic gas in the near field of the metasurface. To understand this point, we remind that the electromagnetic response of the metasurface is composed of evanescent surface plasmonic modes, and a propagating mode. The latter is present in the far field region. The formers are transversally stationary waves with wave number, in the transverse directions, larger than the incident wave. Hence, atoms moving in close proximity to the surface, are coupled to the surface plasmonic modes, but they experience large frequency shift due to Doppler effect. Thus, they will not contribute to the narrow frequency lineshapes as shown in Figures 2(c) and (d) [30].

The spectra measured on both bare fiber and fiber with metasurface have a similar fitted atomic linewidth $\Gamma$ of $2\pi \times 45(5)$ MHz, which is larger than the atomic bare linewidth of $2\pi \times 5.2$ MHz. The Casimir-Polder counts for $\sim \text{Im}\{\Delta C_3\}k^3 = 2$ MHz, which cannot explain the linewidth broadening. We related the linewidth broadening to a transit time broadening of the moving atoms in the optical intensity profile at the fiber tip. Because of the multimode nature of the optical fiber, the optical intensity spatial distribution at the fiber tip looks disordered like a speckle pattern [inset of Figure 3]. We evaluated the characteristic size of the speckle-like grain by computing the autocorrelation function, $C(\mu) = \langle \Delta I(x). \Delta I(x+\mu) \rangle / \langle \Delta I(x) \rangle^2$ of the intensity profile and found a characteristic 1/e width of $\Delta\mu = 4.4 \pm 0.2$ µm, see Figure 3. $\Delta I = I - \langle I \rangle$, where $I(x)$ is the light intensity spatial distribution. $\langle \cdot \rangle$ denotes the 1D spatial averaging. For atoms with thermal velocity of $\bar{v} = 140$ m/s (temperature of 50 °C), the corresponding transit time broadening is $\bar{v}/\Delta\mu = 2\pi \times 32 \pm 2$ MHz, which accounts as the dominant broadening contribution to the $2\pi \times 45$ MHz linewidth. Transit time broadening can be also reformulated in terms of Doppler broadening. To do so, we calculate the surface wavevector distribution at the fiber tip from Figure 3 and further observe that the wavevector distribution has a characteristic width of $\Delta k = 2\pi \times 0.23 \pm 0.09$ µm$^{-1}$. We then compute a residual Doppler broadening [31] of $\Delta k \bar{v} = 2\pi \times 31 \pm 12$ MHz, in agreement with our transit time broadening results. We note also that the dominant Doppler broadening contribution to the atom-light excitation on the fiber tip is given by an effective $NA_{eff} \sim \Delta k/k = 0.20 \pm 0.01$, close to the rated NA of the multimode fiber (NA=0.22). Other sources of broadening such as atom-atom collision have been reported in similar studies [22,27,32] and might play a role, but minor, in our experiment.

In conclusion, we demonstrate the modification of the reflection signal from an atomic vapor and Casimir-Polder potential tuning by coating a metasurface on a fiber tip. This paves the way for fiber-integrated surface engineering of hybrid atoms-metasurface devices. As the light polarization is not maintained in the fiber, we use a two-dimensional array of rings, which exhibit light polarization-independent plasmonic resonances for excitation near normal incidence. Polarization dependence in the atom-plasmon coupling can be restored, for example, if the atomic vapor becomes optically active by applying a magnetic field. In this case, extra tuning of the atom-plasmon Fano resonance can be implemented [33]. The field intensity distribution at the output of the fiber tip has a disordered structure similar to a speckle pattern.

The transport of matter wave in such disordered optical potentials has recently attracted a lot of attention both for non-interacting system leading to Anderson localization [34-36], and for interacting atoms exploring thermalization in presence of many-body interactions, for a recent review see Ref. [37]. Similar studies on the multimode fiber speckle patterns could be done, at which it would be interesting to add a surface plasmon field to gain larger amplitude and shorter length scale of the electromagnetic field. In addition, a single emitter source with a narrow spectrum generated by cold atoms at the nearfield of a fiber nanotip has been proposed [38]. The performance of the single emitter source depends on the collective damping and Purcell effect at the vicinity of the fiber nanotip. In this case, it would be desirable to tailor the fiber nanotip nearfields with plasmonic metamaterials to enhance the collective damping and Purcell effect.

Acknowledgements:

This work is supported by the Singapore Ministry of Education Academic Research Fund Tier3 Grants No. MOE2016-T3-1-006(S), the EPSRC UK Project No. EP/M009122/1 and by Singapore A*STAR QTE program (SERCA1685b0005).

Data availability:

All data supporting this study are openly available from the Nanyang Technological University repository at https://doi.org/10.21979/N9/VABS1C.

Published paper availability:

The published paper is available at Applied Physics Letters **116** (18), 183101 (2020). DOI: 10.1063/1.5142411

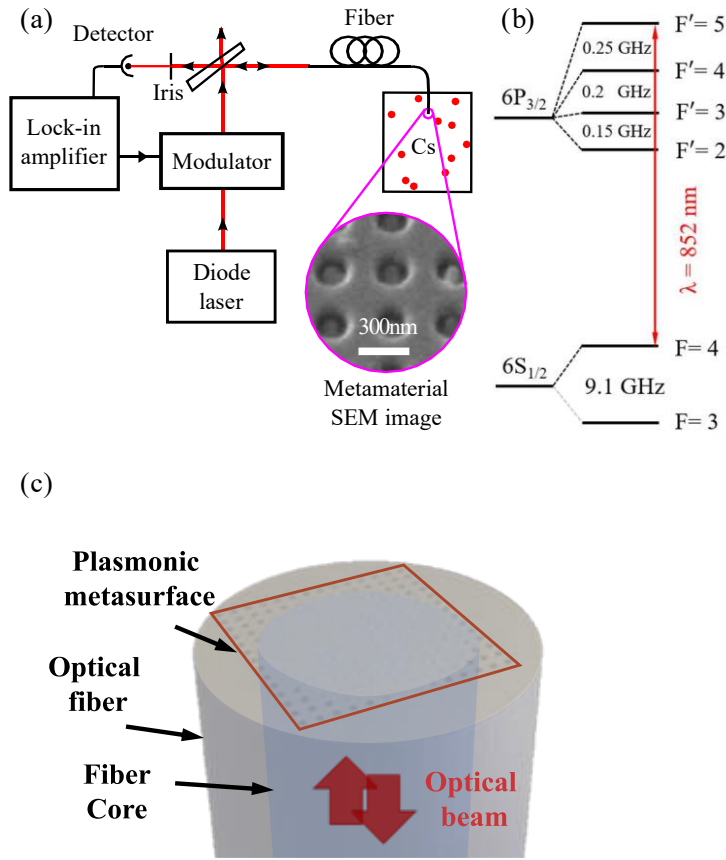

Figure 1. (a) Experimental set-up. Phase modulation spectroscopy measurements are conducted on the $F = 4 - F' = 5$ hyperfine structure of the $6^2S_{1/2} - 6^2P_{3/2}$ D2 in Cesium, see energy diagram in (b). (c) Artistic impression of the optical fiber with plasmonic metasurface (brown box) at the fiber tip immersed into cesium atomic vapor. The light blue area represents the core of the fiber. The red arrows represent the optical excitation and reflection.

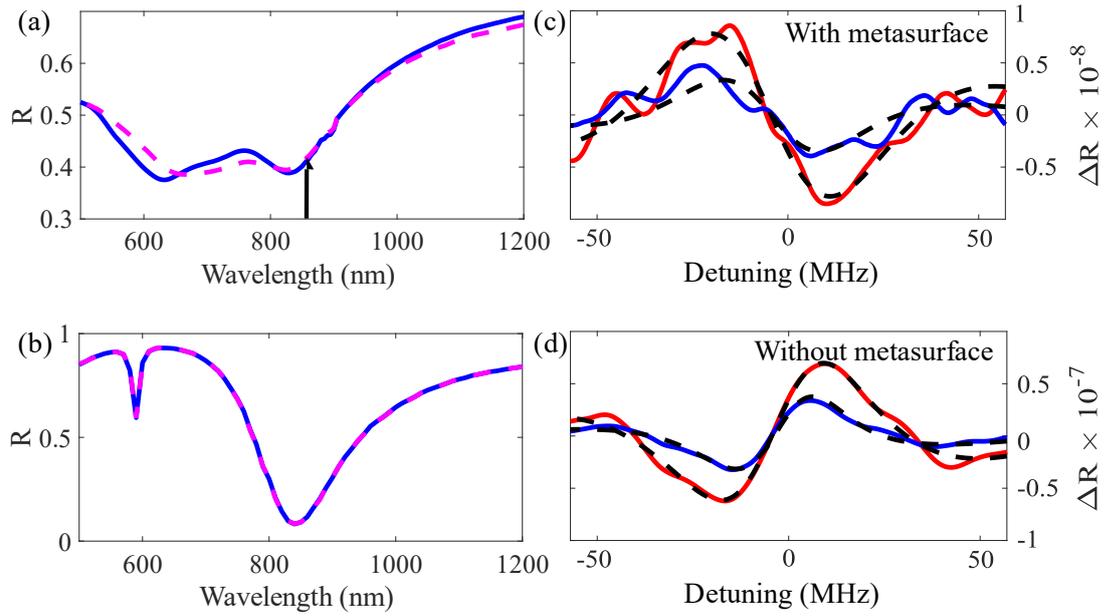

Figure 2. (a) Reflectance spectrum of the plasmonic metasurface for horizontal polarization (resp. vertical) in blue solid curve (resp. dashed magenta). Those spectra are obtained using a commercial microspectrometer with white light, and NA = 0.7. The cesium $D_2$ atomic resonance line (852 nm) is indicated with a black vertical arrow. (b) Comsol simulated reflectance spectrum of the plasmonic metasurface for horizontal polarization (resp. vertical) in blue solid curve (resp. dashed magenta). Identical position of the plasmonic resonance at 830 nm is observed on both measurement and simulation. The dampened resonance of the measurement could be attributed to fabrication imperfections, extra losses and angle dependence of the plasmonic resonance position. (c) & (d). The in-phase and in-quadrature (90° out-of-phase) of the demodulated reflectance spectrum correspond to the red and blue curves, respectively, for fiber with metasurface (c) and bare fiber (d). The exposure time to Cesium is about 3 days. The origin of the detuning corresponds to the strongest $F = 4 - F' = 5$ hyperfine transition of the $D_2$ line. The dashed black curves are line shape fits including the Casimir-Polder interaction model.

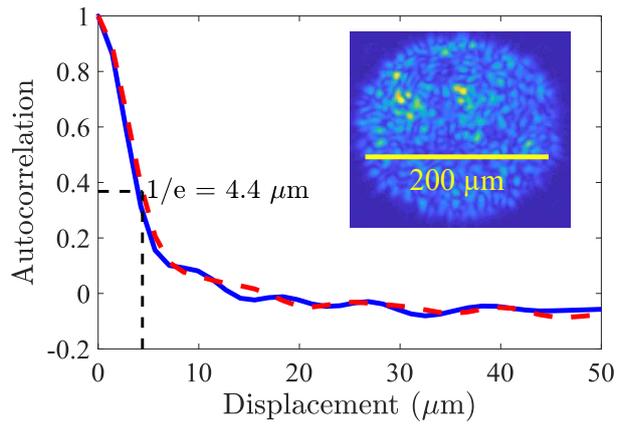

Figure 3. Autocorrelation functions of the intensity at the fiber tip. The solid blue (resp. dashed red) curve corresponds to the horizontal (resp. vertical) direction. Those curves are calculated using the image of the intensity patterns on the fiber tip (in the inset). The autocorrelation functions are characterized by a short correlation range as for a speckle pattern.